\documentclass[aps,prl,twocolumn,groupedaddress]{revtex4}

\usepackage[dvips]{color}
\usepackage{graphicx,color}
\begin{document}

\title{Majorana Fermion Induced Resonant Andreev Reflection }

\author{ K. T. Law$^{1,2}$, Patrick A. Lee$^{2}$, and T. K. Ng$^3$}

\affiliation{$^1$ Institute for Advanced Study, Hong Kong University of Science and Technology, Hong Kong, People's Republic of China\\
$^2$Department of Physics, Massachusetts Institute of Technology, Cambridge, Massachusetts, 02139, USA\\
$^3$Department of Physics, Hong Kong University of Science and Technology, Hong Kong, People's Republic of China}

\begin{abstract}
We describe experimental signatures of Majorana fermion edge states, which form at the interface between a superconductor and the surface of a topological insulator. If a lead couples to the Majorana fermions through electron tunneling, the Majorana fermions induce \textit{resonant} Andreev reflections from the lead to the grounded superconductor. The linear tunneling conductance is $0 $ ($2 e^2/h$) if there is an even (odd) number of vortices in the superconductor. Similar resonance occurs for tunneling into the zero mode in the vortex core. We also study the current and noise of a two-lead device.
\end{abstract}


\maketitle

\emph{Introduction}---Verifying the existence of Majorana fermions in condensed matter systems is an important topic in recent years, because of their potential application for quantum computations which are free from decoherence.\cite{Kitaev,NSSFD} Quantum Hall states as well as superconductors and superfluids with $p_x + i p_y $ pairing symmetry are candidates which support Majorana fermions.\cite{NSSFD,MR,RG,Ivanov,GRA,DNT,FFN,SC} However, Majorana fermions in those systems are yet to be found.

Recently, Fu and Kane \cite{FKPRL08} proposed that Majorana fermions can be created in the vortices of s-wave superconductors deposited on the surface of a three-dimensional topological insulator. \cite{FKM,FKPRB07,MB,ZLQDFZ,Hsieh,Xia} Moreover, chiral Majorana fermion edge states can be created at the interface between a superconductor and the area gapped by ferromagnetic materials \cite{FKPRL08} and several experiments with rather complex geometry have been proposed to study them.\cite{ANB,FKPRL09} In this work, we propose experiments with relatively simple geometry to probe the chiral Majorana fermion edge states. 

More specifically, we study the tunneling current and noise from non-interacting Fermi leads to a grounded superconductor which possesses chiral Majorana edge states at its boundary. The experimental setup is shown in Fig. 1. An s-wave superconducting island is deposited on the surface of a topological insulator whose surface state is described by gapless Dirac fermions.\cite{FKM,FKPRB07} The area outside the superconductor is gapped by ferromagnetic materials. At the interface between the superconductor and the ferromagnetic material, there are gapless chiral Majorana fermion modes surrounding the superconductor.\cite{FKPRL08,ANB,FKPRL09} One or two non-interacting Fermi leads are coupled to this chiral Majorana modes at points $a $ and $b $ with amplitudes $t_1 $ and $t_2$ respectively.

\begin{figure}
\includegraphics[width=2.8in]{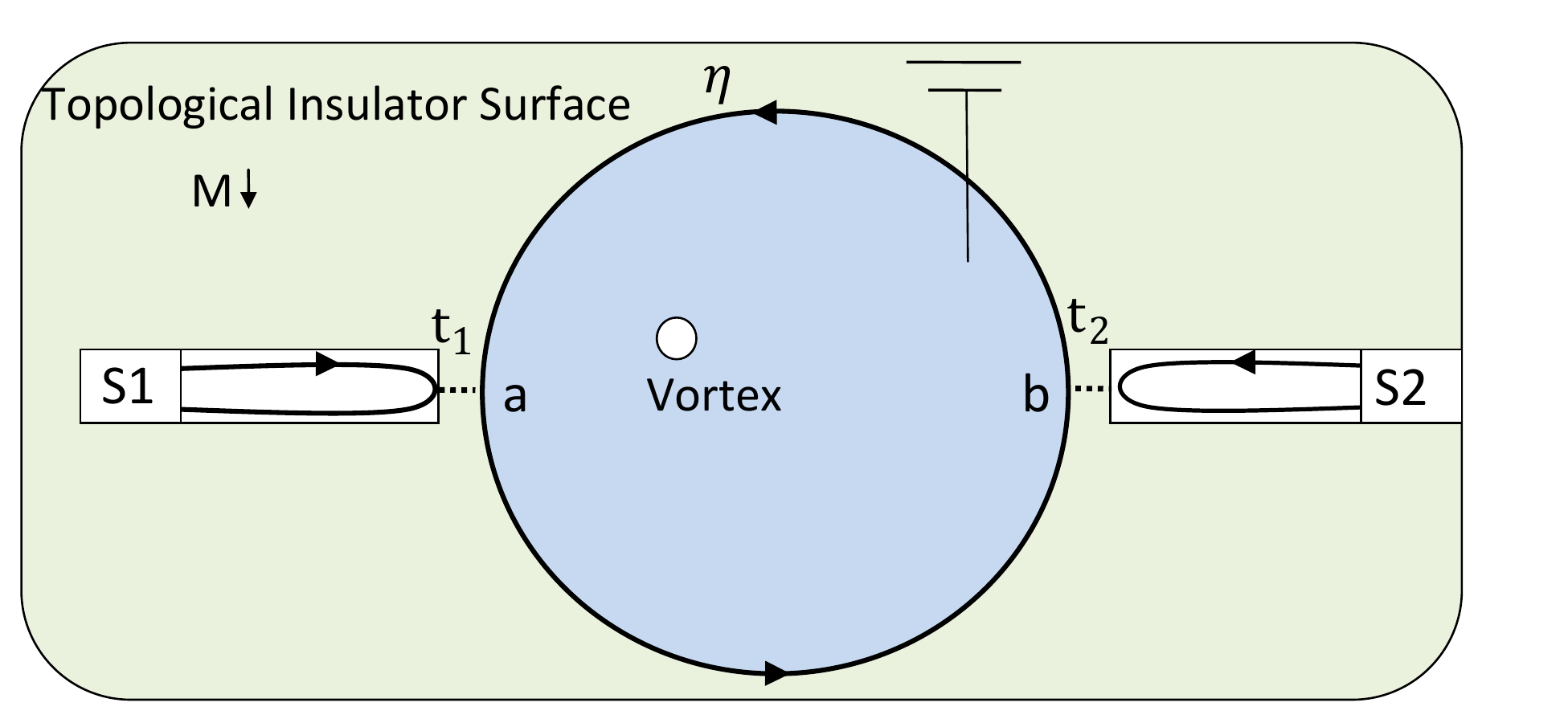}
\caption{\label{f1} A superconducting island is deposited on the surface of a three-dimensional topological insulator. The area outside the superconductor is gapped by ferromagnetic materials. At the interface between the superconductor and the ferromagnetic material, there is a branch of chiral Majorana fermions denoted by $\eta$. Two non-interacting leads are coupled to the Majorana fermions at point $a$ and point $b$ with amplitudes $t_1$ and $t_2$ respectively. The superconductor is grounded.}
\end{figure}

We first consider the single-lead case by setting $t_2$ to zero in Fig. 1. We show that Majorana fermions induce \textit{resonant} Andreev reflections from the lead to the grounded superconductor and result in highly nonlinear I-V curve which is a set of steps. At small voltage, the conductance from the lead to the superconductor is $0$. However, the presence of a vortex in the superconductor changes the conductance to $2e^2/h$. For a two-lead device with non-zero $t_2 $, crossed Andreev reflections may happen. In the small voltage regime, crossed Andreev reflections dominate over local Andreev reflections and the cross current-current correlations of the two leads are maximally positively-correlated. On the other hand, the presence of a vortex in the superconductor in the small voltage regime increases the current dramatically and the cross current-current correlations become maximally negatively-correlated.

\emph{Single-lead Device}---This two-terminal device is shown in Fig. 1 by setting $t_2$ to zero. For simplicity, let us assume that there is only a single mode in lead 1. The Hamiltonian of lead 1 is:
\begin{equation}
H_{L1}= -iv_{f}\sum_{\epsilon= R, L}\sum_{ \sigma=\uparrow,\downarrow} \int_{0}^{+\infty}\psi_{1\epsilon \sigma }^{\dagger}(x)\partial_{x}\psi_{1\epsilon \sigma}(x) dx,
\end{equation}
where the tip of the lead is located at $x=0$ and $ v_{f} $ denotes the Fermi velocity. $ H_{L1}$ contains both left and right moving fields but setting $ \psi_{1L\sigma}(x)=\psi_{1R\sigma}(-x) $ for $x>0 $ maps the Hamiltonian into one described by chiral fields only.
\begin{equation}
H_{L1}=-iv_{f}\sum_{\sigma=\uparrow,\downarrow} \int_{-\infty}^{+\infty}\psi_{1 \sigma}^{\dagger}(x)\partial_{x}\psi_{1 \sigma}(x)dx.
\end{equation}
Let $H_{M0}$ describe the chiral Majorana fermion mode surrounding the superconductor\cite{FKPRL09} and $ H_{T1} $ the coupling term between the lead and the chiral Majorana fermion mode with coupling strength $ t_{1} $. 
The total Hamiltonian is:
\begin{equation}
H_{1}= H_{L1} + H_{M0} + H_{T1},
\end{equation}
where
\begin{equation}
\begin{array}{l}
H_{M0}=iv_{m}\int_{0}^{L}\eta(x)\partial_{x}\eta(x)dx, \qquad \text{and} \\
H_{T1}= - i\frac{1}{\sqrt{2}}t_{1} \sum_{\sigma=\uparrow,\downarrow} \eta(a)[\xi_{\sigma}\psi_{1 \sigma}(0)+\xi_{\sigma}^{*}\psi_{1 \sigma}^{\dagger}(0)].    \label{H2}
\end{array}
\end{equation}
In Eq.\ref{H2}, $v_{m}$ denotes the Fermi velocity of the Majorana mode. It is important to note that the Majorana mode on the surface of the topological insulator includes both spin components and has four instead of two components as in the case of spin polarized p-wave superconductor. Thus, the Majorana mode in Eq.\ref{H2} couples to both spin up and spin down electrons. $\xi_{\sigma}$ are complex numbers with $|\xi_{\sigma}|=1$.

We define two Fermi fields: $ \psi_{1}=\frac{1}{\sqrt{2}}(\xi_{\uparrow}\psi_{1\uparrow}+\xi_{\downarrow}\psi_{1 \downarrow})$ and $ \psi'_{1}=\frac{1}{\sqrt{2}}(\xi_{\uparrow}\psi_{1\uparrow}-\xi_{\downarrow}\psi_{1 \downarrow})$ such that the Majorana fermion couples to $ \psi_{1} $ only.  Dropping $\psi'_{1} $ we have: $H'_{1}= H'_{L1}+H_{M0}+H_{T1}$ where $H'_{L1}=-iv_f\int_{-\infty}^{+\infty}\psi_{1}^{\dagger}(x)\partial_{x} \psi_{1}(x)dx $ and $H_{T1}=-it_{1}\eta(a)[\psi_1 + \psi_{1}^{\dagger}$].

In order to calculate the tunneling current from the lead to the grounded superconductor through its chiral Majorana fermion modes, we first calculate the scattering matrix of the model described by $H'_1$.  Denoting the incoming and outgoing scattering states of the electrons and holes by $ \psi_{1k}(0\mp) $ and $ \psi_{1-k}^{\dagger}(0\mp) $ respectively, the scattering matrix can be written as:

\begin{equation}
\left(
\begin{array}{c}
\psi_{1k}(0+)\\ \psi_{1-k}^{\dagger}(0+)
\end{array}
\right)
=S
\left(
\begin{array}{c}
\psi_{1k}(0-)\\ \psi_{1-k}^{\dagger}(0-)
\end{array}
\right)
\end{equation}  

where
\begin{equation}
S=\left(
\begin{array}{cc}
s^{ee} & s^{eh} \\
s^{he} & s^{hh} \\
\end{array}
\right) =
\frac{1}{Z}
\left(
\begin{array}{cc}
i \sin (\theta/2) & -\tilde{t}_{1}^{2}\cos (\theta/2) \\
-\tilde{t}_{1}^{2}\cos (\theta/2) & i \sin (\theta/2) \\
\end{array}
\right).   \label{SM}
\end{equation} 
In Eq.\ref{SM}, $Z=i \sin (\theta(k,n)/2) + \tilde{t}_{1}^2 \cos (\theta(k,n)/2) $ and $ \theta(k,n)=kL+\pi+n\pi $ is the phase a Majorana fermion with wave vector $k$ acquires when it makes a complete circle around the superconducting island. $L$ is the circumference of the island and $\pi$ is Berry phase contribution from the spin. $ n $ is the number of vortices in the superconductor. $\tilde{t}_1 = t_{1}/(2\sqrt{v_{m} v_{f}})$ is dimensionless.

From the structure of the scattering matrix, it is obvious that when $ \theta(k,0)/2 = kL+ \pi = 2m\pi$ with integer $m$, $ |s^{he}|^2=1 $. In other words, when an incoming electron has energy which matches an energy level of the quantized chiral Majorana modes, there is a resonant Andreev reflection. This means that an incident electron from the lead is converted into a backscattered hole with probability of unity, independent of the coupling strength. We call this Majorana fermion induced resonant Andreev reflection (MIRAR). This is in sharp contrast to the usual normal metal-insulator-superconductor junction in which the Andreev reflection amplitude at fixed subgap energy always decreases with decreasing coupling strength.

To acquire a physical picture of MIRAR, we note that for conventional resonant tunneling, unity transmission requires tuning the coupling strengths of the resonant level with the two leads to equal amplitude (Fig.2a). However, when a lead is coupled to a Majorana fermion mode, the lead plays the role of both an electron lead and a hole lead. Because of its self-Hermitian property, the Majorana fermion is ensured to couple to the electron and hole leads with equal amplitude as can be seen in Eq.\ref{H2}. This results in resonant tunneling from the electron lead to the hole lead (Fig.2b). 

The physical picture shows that MIRAR is a very general phenomenon which happens whenever a discrete Majorana state is coupled to a Fermi lead.  Another interesting example is the tunneling into the Majorana zero mode expected to exist in the vortex core.  This experiment can be performed using a STM tip.  The predicted linear conductance of $2e^2/h$ should be a spectacular signature of the zero mode. When two vortices approach each other, the zero modes are coupled and split into occupied and unoccupied fermions and the resonance disappears once the energy splitting exceeds the voltage and the temperature.  The mathematics is similar to tunneling into two localized Majorana states considered by Nilsson {\em et al.},\cite{NAB} even though they focus on the regime of large splitting and did not consider the resonant tunneling into an isolated vortex.

\begin{figure}
\includegraphics[width=2.8in]{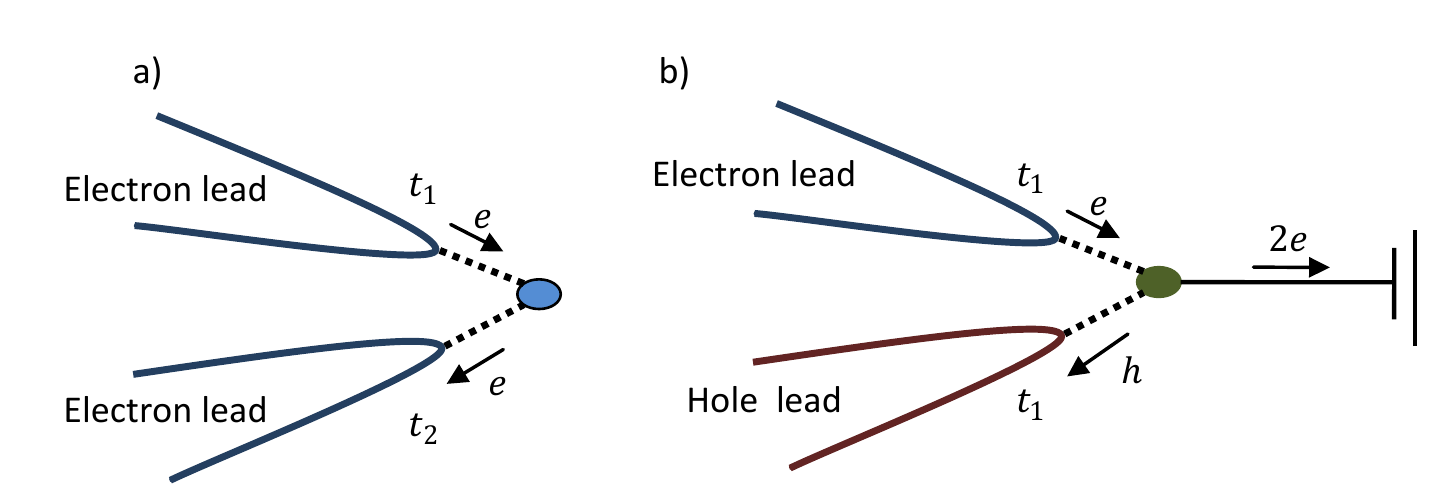}
\caption{\label{f2} a) Conventional resonant tunneling: two leads are coupled to a resonant level with coupling amplitudes $t_1$ and $t_2$ respectively. Resonant tunneling with unity transmission probability can happen only if $t_1=t_2$. b) MIRAR: a single lead coupled to a Majorana level plays the role of both an electron lead and a hole lead.  The coupling amplitudes of the leads to the Majorana level are ensured to be the same. The Majorana mode is attached to a superconductor which is grounded. }
\end{figure}

From the scattering matrix Eq.\ref{SM}, the tunneling current from the lead to the superconductor is:
\begin{equation}
I = \frac{2e}{h}\int_{0}^{eV}T(E)dE=\frac{2e}{h}\int_{0}^{eV}\frac{\tilde{t}_1^{4}(1+\cos\theta)}{1-\cos\theta + \tilde{t}_1^{4}(1+\cos \theta)}dE. \label{current1}
\end{equation}
where $T(E)$ denotes the Andreev reflection probability $|s^{he}|^2$ at energy $E$. Near $\cos\theta = 1$, $T(E)$ can be cast into the resonance form:
$
T(E)=\frac{\gamma_{1}^2}{(E-E_{l})^2 + \gamma_{1}^2},
$
where $ \gamma_1= 2\tilde{t}_1^2 v_{m} \hbar /L $ and 
$E_{l} = (2\ell +n+1)\hbar v_m \pi/L$ 
denote the quantized energy levels of the chiral Majorana fermion modes. 
Provided $\gamma_1$ is less than the level spacing, the differential conductance $dI/dV$ verses $eV$ peaks at $2e^2/h$ whenever the electron energy is resonant with the Majorana mode energy $E_\ell$.  As seen from Fig.3 the resonance is shifted in voltage by half the level spacing when a vortex is added.  For $eV \ll \gamma$, the conductance jumps between $2e^2/h$ and near zero.  We consider this a clear signature of the Majorana mode.

\begin{figure}
\includegraphics[width=2in]{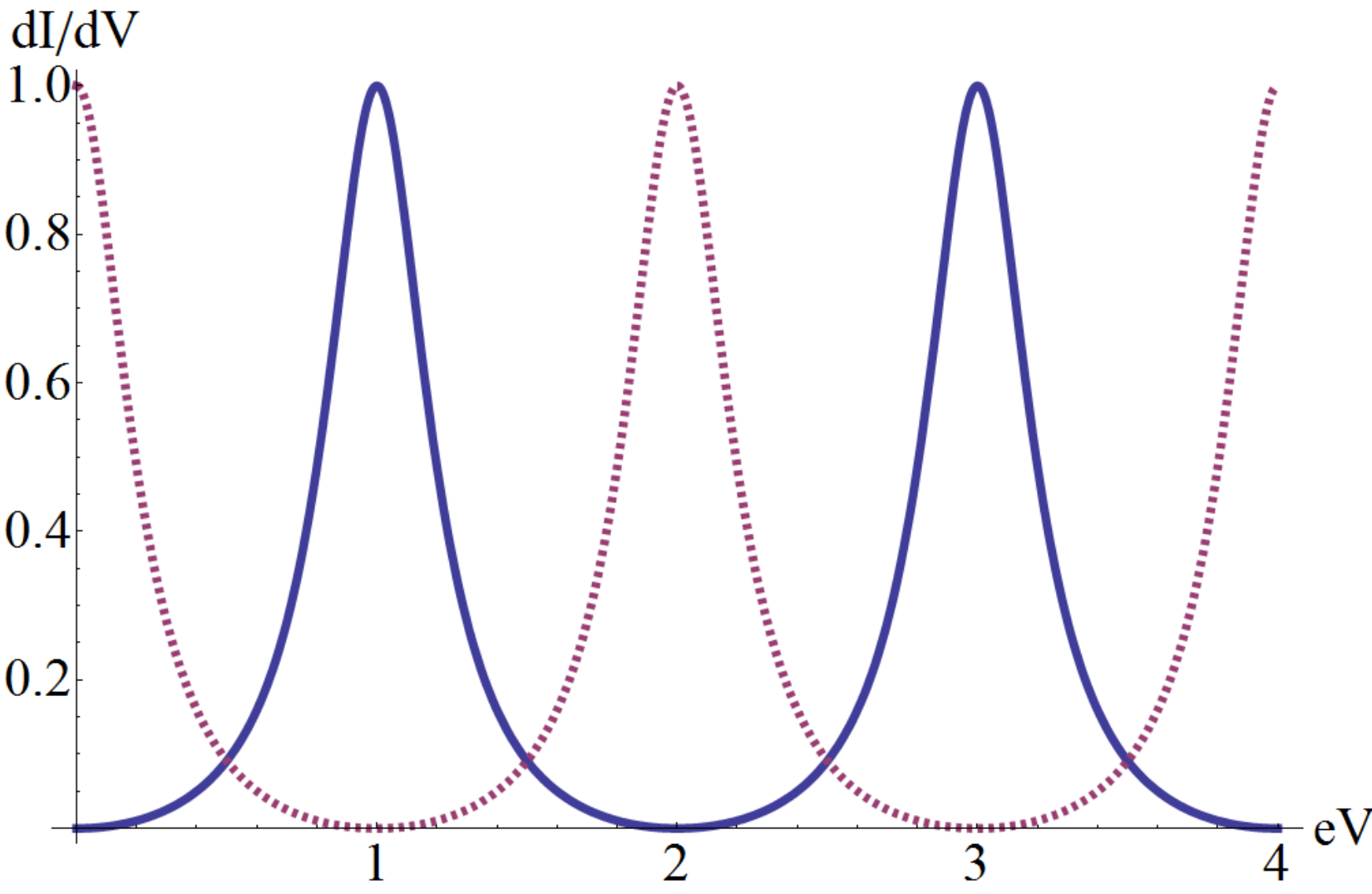}
\caption{\label{f3} $dI/dV$ verses $eV$ with $\tilde{t}_{1}^4 =0.1 $. $eV$ is in units of $\pi \hbar v_{m}/L $ and $dI/dV$ is in units of $\frac{2e^2}{h}$. Solid (dashed) line represents the case with even (odd) number of vortices in the superconductor. }
\end{figure}

\emph{Two-lead Device}--- 
Next we couple to the chiral Majorana mode with one more lead, by setting $t_2 $ to non-zero in Fig. 1.
The new Hamiltonian becomes:
$H_2^\prime = H_1^\prime + H_{L2}^\prime + H_{T2}$, where $H_{L2}^\prime$ and $H_{T2}$ are $H_{L1}^\prime$ and $H_{T1}$ with $\psi_1$, $t_1$ and $a$ replaced by $\psi_2$, $t_2$ and $b$ respectively.

The scattering matrix in the basics of $ (\psi_{1k}, \psi_{2k}, \psi_{1-k}^{\dagger} , \psi_{2-k}^{\dagger} )$ can be written as:
\begin{equation}
S=
\left(
\begin{array}{cc}
1+A & A \\
A & 1+A \\
\end{array}
\right),
\end{equation}
where 
\begin{equation}
A=\frac{1}{Z'}
\left(
\begin{array}{cc}
i\tilde{t}_{1}^{2}\tilde{t}_{2}^2\sin (\frac{\theta}{2}) - \tilde{t}_{1}^2 \cos(\frac{\theta}{2}) & \frac{-2\beta}{1+\alpha \beta}\tilde{t}_{1}\tilde{t}_{2} \cos(\frac{\theta}{2}) \\
\frac{-2\alpha}{1+\alpha \beta}\tilde{t}_{1}\tilde{t}_{2} \cos(\frac{\theta}{2})& i\tilde{t}_{1}^{2}\tilde{t}_{2}^2\sin (\frac{\theta}{2}) - \tilde{t}_{2}^2 \cos(\frac{\theta}{2}) \\
\end{array}
\right).
\end{equation}

In the above equation, $ Z'= -i (1+ \tilde{t}_{1}^{2}\tilde{t}_{2}^2) \sin (\theta/2) + (\tilde{t}_{1}^2 + \tilde{t}_{2}^2) \cos(\theta/2)$, $\tilde{t_i}=t_{i}/(2\sqrt{v_{f} v_{m}})$. $ \theta(k,n)=kL+\pi+n\pi $ is the same as in the two terminal case. $\alpha(\beta)$ is the phase factor acquired by a Majorana mode propagating from point $a(b)$ to point $b(a)$.

Because of the special form of the scattering matrix, the average current from lead $i $ to the grounded superconductor $\bar{I_{i}}$, and the current noise correlators $ P_{ij}$ can be written in a compact form:\cite{NAB}
\begin{equation}
\bar{I_{i}} = \frac{2e}{h}\int_{0}^{eV} (AA^{\dagger})_{ii}dE,
\end{equation}
\begin{equation}
P_{ij}=e\bar{I}_{i}\delta_{ij}+\frac{2e^2}{h}\int_{0}^{eV}[|A_{ij}+(AA^{\dagger})_{ij}|^2 - |(AA^{\dagger})_{ij}|^2]dE,
\end{equation}
where the current noise correlators are defined as 
\begin{equation}
P_{ij}=\int_{-\infty}^{+\infty}<[I_{i}(0)-\bar{I}_{i}][I_{j}(t)-\bar{I}_{j}]> .
\end{equation}
The total current from the leads to the superconductor $\bar{I}=\bar{I}_{1}+\bar{I}_{2}$ is
\begin{equation}
\bar{I}=\frac{2e}{h}\int_{0}^{eV}T(E)dE,
\end{equation}
where
\begin{equation}
T(E)=1-\frac{(1-\tilde{t}_{1}^4 \tilde{t}_2^4)\sin^2(\theta/2)}{(1+\tilde{t}_{1}^2 \tilde{t}_2^2)^2 \sin^2{\theta/2}+(\tilde{t}_{1}^2 +\tilde{t}_2^2)^2 \cos^2(\theta/2)}.
\end{equation}
For $t_1, t_2  <1 $, the resonant Andreev reflection condition is $E-E_{l} << 2\hbar v_f  (\tilde{t}_1^2+\tilde{t}_2^2)/L$, where $E_{l}$ are the energy levels of the Majorana modes. As in the two-terminal device, the I-V curve of the total current is highly non-linear and is a set of steps. In the small voltage regime with $ eV << 2\hbar v_f  (\tilde{t}_1^2+\tilde{t}_2^2)/L$, the conductance is  $\frac{2\tilde{t}_1^2 \tilde{t}_2^2}{1+\tilde{t}_1^2 \tilde{t}_1^2}\frac{2e^2}{h} $ when there are no vortices in the superconductor. It is important to note that the linear conductance is not $0$ as  in the two-terminal case. We argue below that this is the consequence of crossed Andreev reflections. If a vortex is created in the superconductor, the conductance becomes $ \frac{2e^2}{h} $ because of resonant Andreev reflection induced by the $E_{l}=0$ Majorana mode.  

Crossed Andreev reflection is a process which an incoming electron from say, lead 1, is turned into an outgoing hole in lead 2. As a result, one electron from each lead tunnels into the superconductor to form a Copper pair. This process is not allowed in the single-lead device. We show below that measuring the shot noise of the tunneling currents can be used to reveal the mechanism of the tunneling processes, whether they are due to local or crossed Andreev reflections. In the following, we analyses the shot noise in the small voltage regime with $ eV << 2\hbar v_f  (\tilde{t}_1^2+\tilde{t}_2^2)/L$.

In this regime and in the absence of vortices, we have:
\begin{equation}
\bar{I}_{1}=\bar{I}_{2}=\bar{I}/2=\frac{2e^2V}{h} \frac{\tilde{t}_1^2 \tilde{t}_2^2}{1+\tilde{t}_1^2 \tilde{t}_2^2}.
\end{equation}
It is important to note that the individual contributions to the total current $ \bar{I} $ from the two leads are equal and depend on the products of the coupling strengths $\tilde{t}_1 \tilde{t}_2$ only. This is a strong indication that the tunneling processes are dominated by crossed Andreev reflections. In order to verify this, we note that the noise correlators are given by:
\begin{equation}
P_{11}=P_{22}=P_{12}=P_{21}=\frac{2e^3V}{h} \frac{\tilde{t}_1^2 \tilde{t}_2^2}{(1+\tilde{t}_1^2 \tilde{t}_2^2)^2} \label{correlator2}
\end{equation}

From Eq.\ref{correlator2}, we have $ P_{ii}= \frac{e}{1+\tilde{t}_1^2 \tilde{t}_2^2} \bar{I}_{i}$. For $ (\tilde{t}_1 \tilde{t}_2)^2 <<1 $, the Fano factor which is defined as $P_{ii}/e\bar{I}_i$ approaches 1 indicating that crossed Andreev reflections dominate in this small voltage regime. 

Another point is that the cross current-current correlation function $ P_{12}=(P_{11}+P_{22})/2 $. As pointed out in Ref.\onlinecite{NAB}, for any stochastic process the cross correlator has to satisfy the condition: $|P_{12}| \le \frac{1}{2}(P_{11}+P_{22}).$ Thus, the currents from the leads are positively correlated in a maximal way, which is a consequence of crossed Andreev reflections and consistent with the finding of Ref.\onlinecite{NAB} 
where two leads are coupled to two localized Majorana modes.

In the small voltage regime with odd number of vortices, the average current from lead $ i $ to the superconductor is $ \bar{I}_{i}= \frac{2e^2V}{h}\frac{\tilde{t}_{i}^2}{\tilde{t}_{1}^2+\tilde{t}_{2}^2}$ and the total current is $ \bar{I}= \frac{2e^2V}{h}$, independent of the coupling strengths. In this regime,
\begin{equation}
P_{11}=P_{22}=-P_{12}=-P_{21}= \frac{\tilde{t}_{1}^2 \tilde{t}_{2}^2 e}{(\tilde{t}_{1}^2+\tilde{t}_{2}^2)^2} \frac{2e^2V}{h}.  \label{CC2}
\end{equation}

A few comments of this result are in order. First, the noise power of each individual leads are non-zero and $P_{11}= \frac{\tilde{t}_{2}^2}{\tilde{t}_{1}^2+\tilde{t}_{2}^2}e \bar{I}_{1} = P_{22}= \frac{\tilde{t}_{1}^2}{\tilde{t}_{1}^2+\tilde{t}_{2}^2}e \bar{I}_{2}$. To understand this result better, let us assume $ \tilde{t}_1 << \tilde{t}_{2}$. In this limit, $ P_{11} \approx e \bar{I}_1 $. In other words, the Fano factor is $1$. Physically, it means that the tunneling events in lead 1 are dominated by crossed Andreev reflections, which give a Fano factor of 1 instead of 2 as in the case of tunneling into a conventional superconductor which is dominated by local Andreev reflections.

On the other hand, the Fano factor at lead 2 approaches zero. This is expected because the majority of the current from the two leads to the superconductor is carried by lead 2. When $ \tilde{t}_1 << \tilde{t}_{2}$ almost all the incoming electrons from lead 2 are locally Andreev reflected. This is analogous to the situation of a ballistic tunneling junction with tunneling probability $T$
where the Fano factor is suppressed by a factor of $(1-T)$. For $ T \approx 1$, the Fano factor approaches zero.

One more interesting point is that the total noise power $ P_{T}=\sum_{i,j=1,2}P_{ij} $ is zero and $P_{12}=-\frac{1}{2}(P_{11}+P_{22})$. In other words, the two currents are \textit{negatively} correlated in a maximal way. It can be argued that this is a result of crossed Andreev reflection and MIRAR.

\emph{Conclusion and Discussion}---We show that Majorana modes which couple to Fermi leads induce resonant Andreev reflections from the leads to the superconductor. At small voltage limit, the conductance from a single lead to the superconductor is $0 $ ($2 e^2/h$) if there is an even (odd) number of vortices in the superconductor. Attaching one more lead to the Majorana fermion edge states introduces crossed Andreev reflection. The currents from the two leads can be maximally positively-correlated or maximally negatively-correlated, depending on the parity of the number of vortices.

Our discussion can easily be extended to finite temperature $T$ and it is clear that the effects we discussed will be smeared out if the temperature is larger than the level spacing of the Majorana fermions,  $T \gtrsim \hbar v_{m}/L$. This puts a rather stringent, but not impossible, requirement on the size of the dot and position of the Fermi level relative to the Dirac node in order to minimize $L$ and maximize $v_{m}$.\cite{FKPRL09} On the other hand, resonant tunneling into the Majorana zero mode in the vortex core is subject to the less stringent condition $T < \Delta E$ where $\Delta E \sim \Delta^2/E_F$ is the level spacing of states in the core.

\emph{Acknowledgments}--- It is our pleasure to thank D. Feldman, L. Fu, N. Nagaosa, Z.D. Wang and especially C. Kane for insightful discussions. KTL is supported by the IAS-HKUST postdoc fellowship, PAL acknowledge the support of NSF DMR0804040 and the hospitality of IAS-HKUST.

\end{document}